# Near-Wall Scaling and Separation Prediction of a Rotation-Based Subgrid-Scale Stress Model

Jiawei Chen, Yifei Yu, Emran Hossen, Chaoqun Liu*

Department of Mathematics, University of Texas at Arlington, Arlington, Texas 76019, USA

* Correspondence: cliu@uta.edu

**Abstract:** This paper presents an in-depth analysis of a novel subgrid-scale stress model proposed in 2022, which utilizes the rotational part of the velocity gradient as the velocity scale for computing eddy viscosity. This study investigates the near-wall asymptotic behavior and separation prediction capability of this model for the first time. Two canonical flows—fully-developed turbulent channel flow and periodic hill flow—are selected for analysis. The eddy viscosity predicted by this model correlates well with the visualized vortices and exhibits an asymptotic behavior of $O(y)$ near the walls. The dimensionless eddy viscosity, like that of the Wall-Adapting Local Eddy Viscosity (WALE) subgrid model, remains within a small numerical range of $10^{-2}$ to $10^{-4}$. The power spectral density results reveal the asymptotic behavior of the velocity scale in the dissipation range, following a $-10/3$ scaling law. Additionally, this model predicts velocity profiles more accurately than the Smagorinsky model, even when using Van Driest damping. For the periodic hill case, this model predicts the reattachment point with only a 6.9% error, compared to 14.0% for the Smagorinsky model and 16.4% for the Smagorinsky model with Van Driest damping. In near-wall regions with separation, this model achieves even greater accuracy in Reynolds stress prediction than the WALE model, demonstrating its superior potential for separated flow simulations.

**Keywords:** subgrid-scale stress model, rotation, eddy viscosity, separation, Reynolds stress

# 1. Introduction

In large eddy simulation (LES), the large-scale motions are directly resolved, while the effects of small-scale motions are modeled. In terms of computational cost, LES falls between Reynolds-averaged Navier-Stokes (RANS) models and direct numerical simulation (DNS)[1]. Since LES directly resolves large-scale unsteady dynamics, it provides a more accurate representation of turbulent flows. It is particularly suitable for flows with significant separation, such as periodic hills, backward-facing steps, and airfoils with ice accretion.

In the LES approach, scales smaller than the grid size are not resolved but accounted for through the subgrid-scale (SGS) stress model. Most SGS stress models are based on an eddy-viscosity assumption to model the SGS tensor. Smagorinsky[2] developed a SGS stress model in which the SGS stress is assumed to be scaled with the local strain rate of the large scales. This model has demonstrated reasonable accuracy in simulations of decaying homogeneous isotropic turbulence[3][4]. However, the SGS stress predicted by this model does not vanish to zero in laminar flows or near-wall regions, rendering it incapable of simulating transition and wall-bounded flows accurately. To address this issue, damping functions[5][6] have been introduced to reduce the SGS stress near walls. Germano[7] developed the dynamic Smagorinsky model, in which the model coefficient is computed from the local flow state rather than a prescribed constant coefficient. The dynamic Smagorinsky model accurately dissipates energy from large scales in isotropic decaying turbulence and vanishes appropriately in laminar and transitional flows[8]. However, regularization procedures like clipping or spatial averaging are often needed to ensure numerical stability[9].

The Wall-Adapting Local Eddy Viscosity (WALE) model[10] is another SGS model that overcomes some limitations of the Smagorinsky model family mentioned above. The WALE model introduces a velocity scale based on a different invariant, which accounts for both strain and rotation effects. It exhibits appropriate near-wall behavior, vanishing to zero while following a scaling law of $O(y^3)$. Similar to the Smagorinsky model, the model coefficient in WALE is not a universal constant and requires case-specific calibration, especially for complex geometries. To address this, Toda et al.[11] applied the Germano–Lilly procedure to the WALE model, resulting in the Dynamic WALE model. However, they observed that the dynamic WALE model overestimated the turbulent eddy viscosity because of a large WALE constant in the near-wall region.

This paper investigates the near-wall scaling and separation prediction capabilities of the Liutex[12]-based SGS model proposed by Ding et al[13]. Hereafter, this model proposed by Ding et al. [13] will be referred to as the "present model". They introduced Liutex[12], a measure of rigid rotation, as the velocity scale for computing eddy viscosity, motivated by the fact that rigid rotation is zero in the viscous sublayer. They initially verified that the eddy viscosity indeed vanishes at the wall in channel flow. However, the near-wall scaling and separation prediction capabilities have not yet been explored. This paper first presents the fluid mechanism underlying this subgrid scale stress model in channel flow. The spectral characteristic, velocity profile, and near-wall scaling are then examined. Furthermore, the classical periodic hill case is employed to assess its performance in predicting large-scale separation. Key flow quantities, including friction velocity, mean velocity profiles, and Reynolds stress, are thoroughly validated. This model is compared against the Smagorinsky model, the Smagorinsky model with Van Driest damping, and the WALE model.

The remainder of this paper is organized as follows. Section 2 introduces the governing equations, subgrid-scale models, and numerical methods. Section 3 validates the model capabilities using the turbulent channel flow, and turbulent periodic hills. Finally, Section 4 provides the conclusions of this study.

## 2. Methodology

### 2.1 Governing equations and subgrid-scale stress models

The governing equations for LES of incompressible flow are obtained by applying grid filter to the Navier-Stokes equations.

$$\frac{\partial \bar{u}_i}{\partial x_i} = 0 \tag{1}$$

$$\frac{\partial \bar{u}_i}{\partial t} + \frac{\partial \bar{u}_j \bar{u}_i}{\partial x_j} = -\frac{1}{\rho}\frac{\partial \bar{p}}{\partial x_i} + \nu \frac{\partial^2 \bar{u}_i}{\partial x_j x_j} - \frac{\partial \tau_{ij}^{sgs}}{\partial x_j} \tag{2}$$

where $\tau_{ij}^{sgs} = \overline{u_i u_j} - \bar{u}_i \bar{u}_j$ is the subgrid stress. The isotropic component of the SGS stress is typically absorbed into the pressure, resulting in a pseudo-pressure field ($p \to p + \rho\tau_{kk}^{sgs}$). SGS closure models based on the eddy viscosity assumption and the Boussinesq hypothesis take the following form:

$$\tau_{ij}^{sgs} - \frac{1}{3}\tau_{kk}\delta_{ij} = -2\nu_t \overline{S}_{ij}, \text{ where } \overline{S}_{ij} = \frac{1}{2}\left(\frac{\partial \overline{u}_i}{\partial x_j} + \frac{\partial \overline{u}_j}{\partial x_i}\right) \quad (3)$$

where $\delta_{ij}$ is the Kronecker delta function and $\overline{S}_{ij}$ is the strain-rate tensor.

## A. Smagorinsky model

The Smagorinsky subgrid scale model was proposed by Joseph Smagorinsky[2] in the 1960s. It is based on the eddy viscosity assumption, which assumes a linear relationship between the SGS shear stress and the rate of the resolved strain tensor $\overline{S}_{ij}$. This model serves as the foundation for many subsequent SGS models. The basic Smagorinsky model is given by

$$\nu_t = (C_s \Delta)^2 |S|, \text{ where } |S| = \sqrt{2\overline{S}_{ij}\overline{S}_{ij}} \quad (4)$$

Where $C_s$ is a model coefficient and $\Delta$ is the subgrid length scale. The subgrid length is defined as $\Delta = \sqrt[3]{\Delta x \Delta y \Delta z}$, where $\Delta x$, $\Delta y$, and $\Delta z$ denote the grid spacings in the *x*-, *y*-, and *z*-directions, respectively. The optimal model coefficient depends on flow patterns. Lilly[14] determined that for isotropic turbulence within the inertial subrange, $C_s \approx 0.17$, while Deardorff[15] proposed $C_s \approx 0.1$ for wall-bounded turbulent shear flows. To account for near-wall behavior, the Van Driest damping function[16] is typically applied and is defined as:

$$D = 1 - \exp\left(-\frac{y^+}{A^+}\right) \tag{5}$$

The final length scale is given by:

$$\Delta = \min\left(\frac{\kappa y}{C_s} D, \Delta_g\right) \tag{6}$$

where $\Delta_g$ is a geometric-based delta function such as the cube-root volume delta.

## B. WALE model

The WALE model[10] is based on the square of the velocity gradient tensor and accounts for both strain and rotation effects on the smallest turbulent scales that are resolved. Moreover, it correctly recovers the $y^3$ near-wall scaling for eddy viscosity without the need for a dynamic procedure. The traceless symmetric part of the square of the velocity gradient tensor ($\bar{g}_{ij} = \partial u_i / \partial x_j$) is given by

$$S_{ij}^d = \frac{1}{2}\left(\bar{g}_{ij}^2 + \bar{g}_{ji}^2\right) - \frac{1}{2}\delta_{ij}\bar{g}_{kk}^2 \tag{7}$$

Where $\bar{g}_{ij}^2 = \bar{g}_{ik}\bar{g}_{kj}$ and $\delta_{ij}$ is the Kronecker symbol.

$$\nu_t = (C_w \Delta)^2 \frac{\left(S_{ij}^d S_{ij}^d\right)^{3/2}}{\left|\bar{S}_{ij}\bar{S}_{ij}\right|^{5/2} + \left(S_{ij}^d S_{ij}^d\right)^{5/4}} \tag{8}$$

Where $C_w$ is the model coefficient and is set to 0.5 for the following computations. In the case of pure shear (where $\bar{g}_{ij} = 0$ except for $\bar{g}_{12} = 0$), $S_{ij}^d S_{ij}^d = 0$, leading to $\nu_t = 0$ in the laminar sublayer near the wall

## C. Present SGS stress model

The model proposed by Ding et al. [13] uses rigid rotation ($\boldsymbol{R}$) as the velocity scale in the eddy viscosity calculation. Rigid rotation, also represented by Liutex[12], was originally used for vortex identification and analysis[17][18][19]. Unlike other vortex identification methods, such as Q[20], $\lambda_2$ [21], $\Delta$ [22] and $\lambda_{ci}$ [23], $\boldsymbol{R}$ is a vector. Its direction represents the axis of rotation, while its magnitude is twice the local angular velocity of rigid rotation. Kolář and Šístek[24] showed that $\boldsymbol{R}$ is unaffected by stretching or shear, confirming its robustness in vortex identification. The explicit formula of $\boldsymbol{R}$[25] vector is given by

$$\boldsymbol{R} = R\boldsymbol{r} = \left[ \boldsymbol{\omega} \cdot \boldsymbol{r} - \sqrt{(\boldsymbol{\omega} \cdot \boldsymbol{r})^2 - 4\lambda_{ci}^2} \right] \boldsymbol{r} \tag{9}$$

Where $\boldsymbol{\omega}$ is the vorticity vector, $\lambda_{ci}$ is the imaginary part of the complex conjugate eigenvalue of the velocity gradient tensor, and $\boldsymbol{r}$ is the eigenvector corresponding to the real eigenvalue.

In this subgrid model, the subgrid length scale remains the same as in the Smagorinsky model, and the eddy viscosity is ultimately given by

$$\nu_t = (C_s \Delta)^2 |\boldsymbol{R}| \tag{10}$$

where $\Delta$ is the grid filter width and $|\boldsymbol{R}|$ is the magnitude of vector $\boldsymbol{R}$. The model coefficient $C_s$ is set to 0.17.

## 2.2. Numerical methods

In this paper, the open-source software OpenFOAM[26] is used, which is based on the finite volume method for discretizing and solving the governing equations. The flow variables are stored at the centroids of the control volumes (CVs), where the values approximate the local properties of the CV.

In the present simulations, the second-order backward differencing scheme was used for time marching. The face-fluxes of momentum were calculated using a linear interpolation scheme. The same scheme was also used to evaluate the values of the gradients in the centroids of the faces. The face-fluxes of the subgrid scale turbulent kinetic energy were calculated using a TVD interpolation scheme based on upwind and central differencing. The scheme is based on a flux limiter of the form, $\max(\min(2r, 1), 0)$, where $r$ is the ratio of successive gradients.

The solver pimpleFoam was used to solve the equations. The algorithm implemented in the solver is based on a blend of the transient SIMPLE and PISO algorithms, a thorough description of which can be found in Ferziger et al.[27] and Versteeg et al.[28]. The convergence criterion for the numerical solutions is that there is an absolute root-mean-square residual of all equations less than $1.0\times10^{-6}$.

## 3. Results and discussions

### 3.1. Channel flow

Fully developed turbulent channel flow is a theoretical prototype characterized by flow between two infinite parallel planes driven by a constant pressure gradient. It is simulated in a domain of $(L_x, L_y, L_z) = (4h, 2h, 2h)$, with $h$ the channel half-height. Non-slip boundary conditions are imposed in the $y$ direction and periodic boundary conditions are applied to the $x$ and $z$ directions. The bulk Reynolds number is defined as $Re_b = 2u_b h/\nu$, with $u_b$ the bulk velocity and $\nu$ the kinematic viscosity. The friction Reynolds number is defined as $Re_\tau = u_\tau h/\nu$, where $u_\tau = (\tau_w/\rho)^{1/2}$ is the friction velocity. The $Re_b = 13350$, which corresponds to $Re_\tau = 395$. In Table 1, three grid resolutions are used for grid convergence assessment, namely the coarse grid (G1), medium grid (G2), and fine grid (G3). $N_x$, $N_y$, and $N_z$ denote the number of grid points in the x-, y-, and z-directions, respectively, while $\Delta x$, $\Delta y$, and $\Delta z$ represent the corresponding grid spacings. The non-dimensional wall-normal grid spacing $\Delta x^+$ decreases from approximately 23.20 to 5.80 with a constant aspect ratio of $AR = \Delta x/\Delta z = 1.5$. Fig. 1(a) shows the computational domain and medium grid. Fig. 1(b) shows the instantaneous velocity field.

Table 1: Computational grids for channel flow

| Mesh sets | ($N_x \times N_y \times N_z$) | $\Delta x^+$ | $\Delta y^+$ | $\Delta z^+$ | Total |
|---|---|---|---|---|---|
| Coarse (G1) | $80 \times 50 \times 60$ | 34.80 | 3.48 | 23.20 | 0.24M |
| Medium (G2) | $160 \times 100 \times 120$ | 17.40 | 1.74 | 11.60 | 1.92M |
| Fine (G3) | $320 \times 200 \times 260$ | 8.70 | 0.87 | 5.80 | 15.36M |

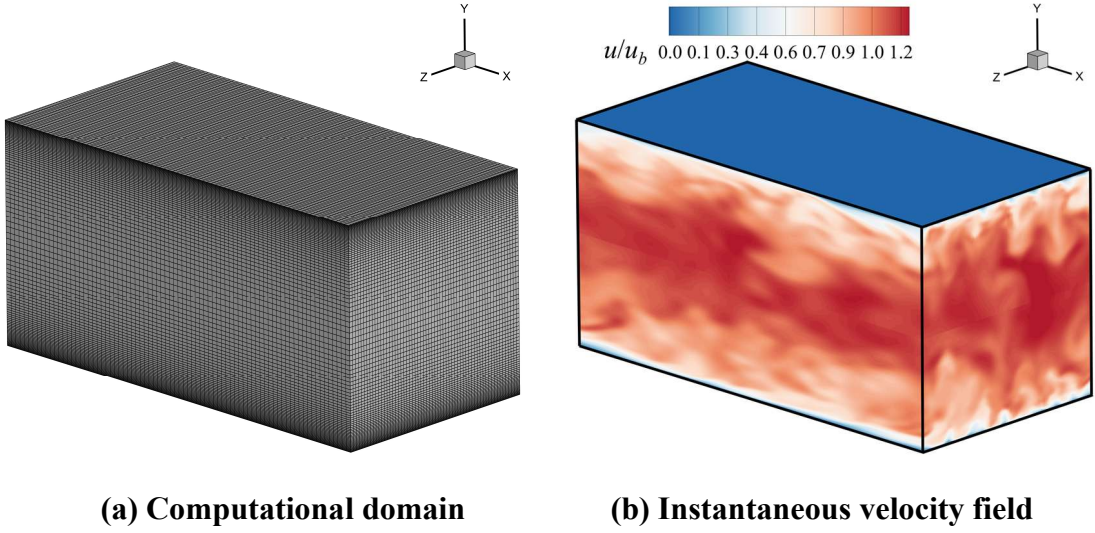

(a) Computational domain　　　(b) Instantaneous velocity field

Fig. 1.  Computational domain and instantaneous flow of channel flow

Fig. 2 presents the instantaneous vortex structures identified using different vortex identification methods (vorticity magnitude, Q-criterion, and $|\boldsymbol{R}|$). Fig. 2(a) shows the vortex structures identified using the vorticity magnitude ($|\boldsymbol{\omega}|$). Numerous vortices can be observed near the wall. However, vorticity-based identification is often severely contaminated by shear, leading to a significant overestimation near the wall. Fig. 2(b) shows the identification results using the Q-criterion, which confirms this issue.

However, the Q-criterion depends on a prespecified threshold, making it difficult to identify weak vortices. The $|\mathbf{R}|$ method (Fig. 3(c)) can effectively eliminate shear contamination (compared to the vorticity-based method) and successfully capture weak vortices (compared to the Q-criterion).

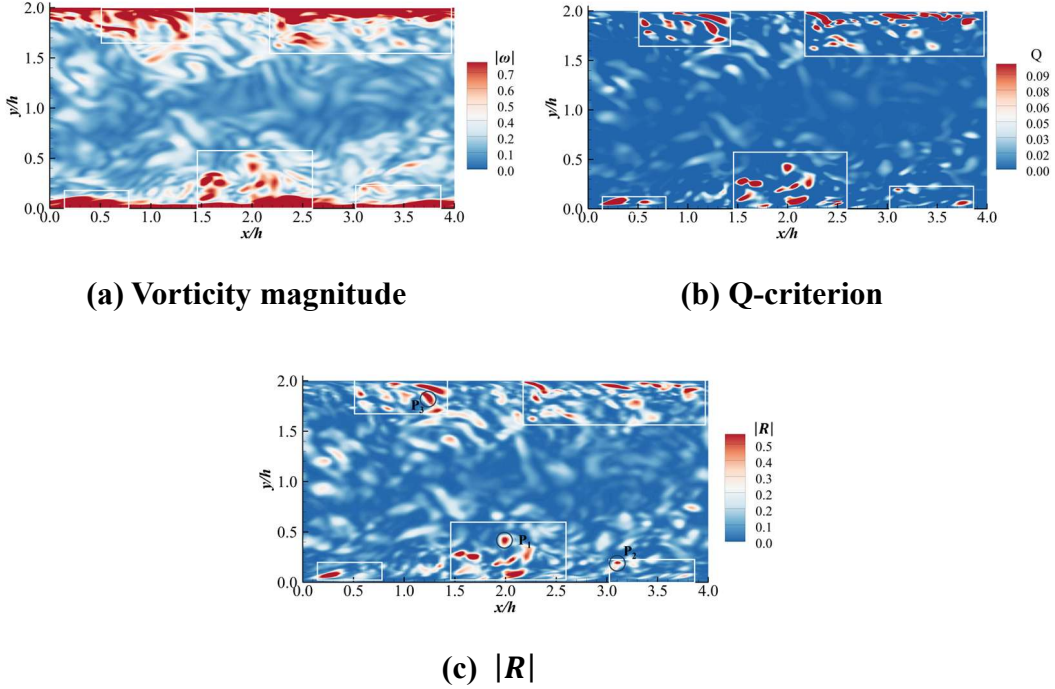

(a) Vorticity magnitude        (b) Q-criterion

(c) $|\mathbf{R}|$

**Fig. 2. Instantaneous vortex structures identified using different vortex identification methods (vorticity magnitude, Q-criterion, and $|\mathbf{R}|$)**

Compared to the Q-criterion vortex identification method, another advantage of the $|\mathbf{R}|$ method is that $\mathbf{R}$ is a vector, allowing it to describe the rotation direction. Fig. 3 presents the three components ($R_x$, $R_y$, $R_z$) of the $\mathbf{R}$ vector. For the three rotating locations in Fig. 2(c), their rotation directions can be analyzed using these components. At position P1, $R_x$ and $R_z$ are stronger than $R_y$, indicating that the rotation axis lies closer to the x-z plane. In fact, the exact rotation direction can be determined from $R_x$, $R_y$ and $R_z$. $\boldsymbol{R}$ itself is a physical quantity, representing twice the local angular velocity. At position P2, $R_z$ is significantly stronger than $R_x$ and $R_y$, indicating that the rotation axis is predominantly aligned with the z-direction. At position P3, $R_y$ is significantly stronger than $R_x$ and $R_z$, suggesting that the rotation axis is closer to the y-direction.

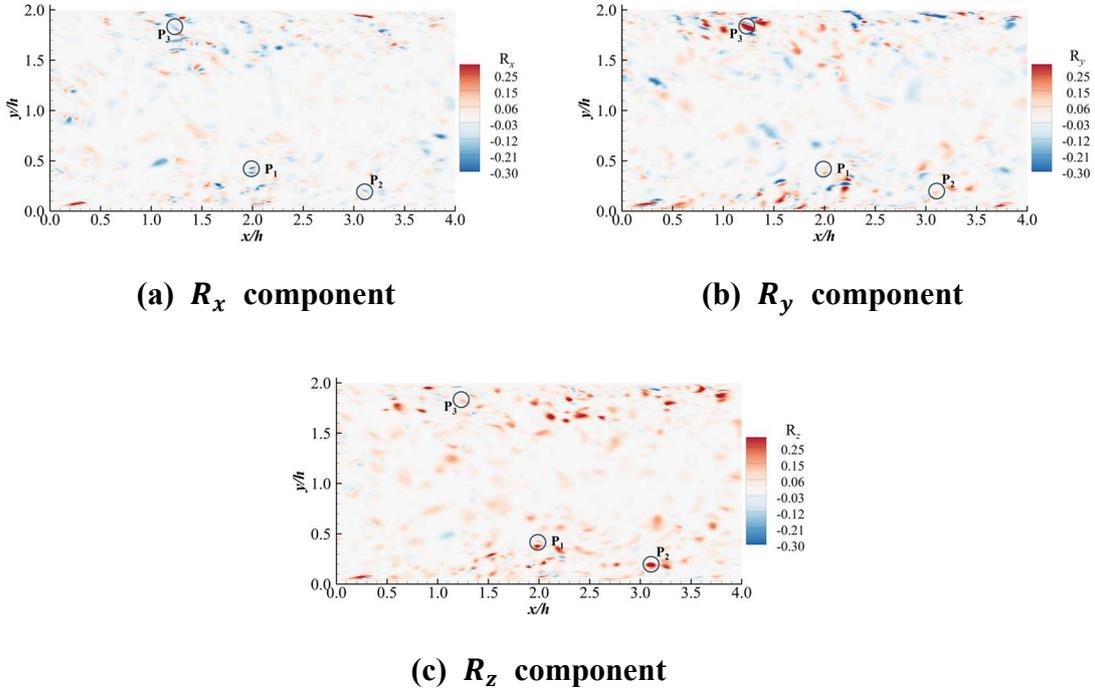

Fig. 3. Three components ($R_x$, $R_y$, $R_z$) of the $\boldsymbol{R}$ vector

Fig. 4 presents the distribution of eddy viscosity $\mu_t$ in the instantaneous flow field. It can be observed that the eddy viscosity distribution corresponds to the $|\boldsymbol{R}|$ distribution in Fig. 2(c). Where $|\boldsymbol{R}|$ exists, eddy viscosity is also present; where $|\boldsymbol{R}|$ is large, eddy viscosity is also large. Therefore, this subgrid stress model is highly consistent with physics, as it is vortex-dependent.

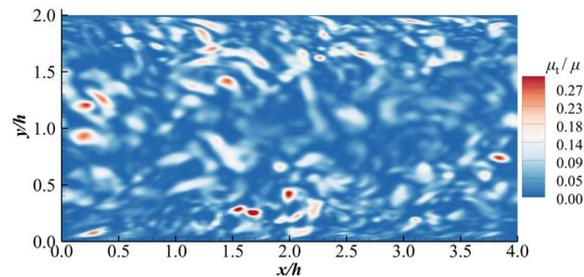

**Fig. 4. Distribution of eddy viscosity $\mu_t/\mu$ in the instantaneous flow field**

Fig. 5 shows the power spectrum density (PSD) of different vortex identification methods at various wall-normal locations ($y^+ =$ 2.5, 17.5 and 70). $y^+ = 2.5$ is in the viscous sublayer, $y^+ = 17.5$ is in the buffer layer, and $y^+ = 70$ is in the log-law layer. In the inertial subrange, all vortex identification methods exhibit Kolmogorov's −5/3 scaling law. Compared to the vorticity and Q-criterion methods, $|R|$ exhibits an asymptotic line in the dissipation range, following a −10/3 slope, which is in accordance with previous findings of Xu et al.[29] and Yan et al.[30].

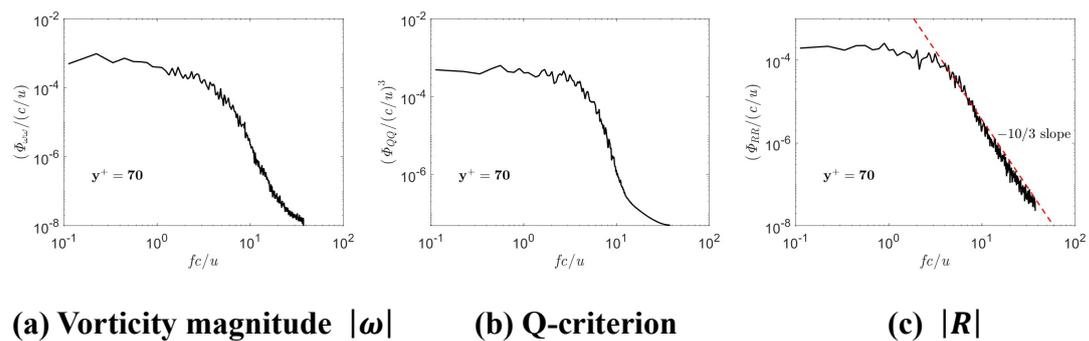

(a) Vorticity magnitude $|\omega|$    (b) Q-criterion    (c) $|R|$

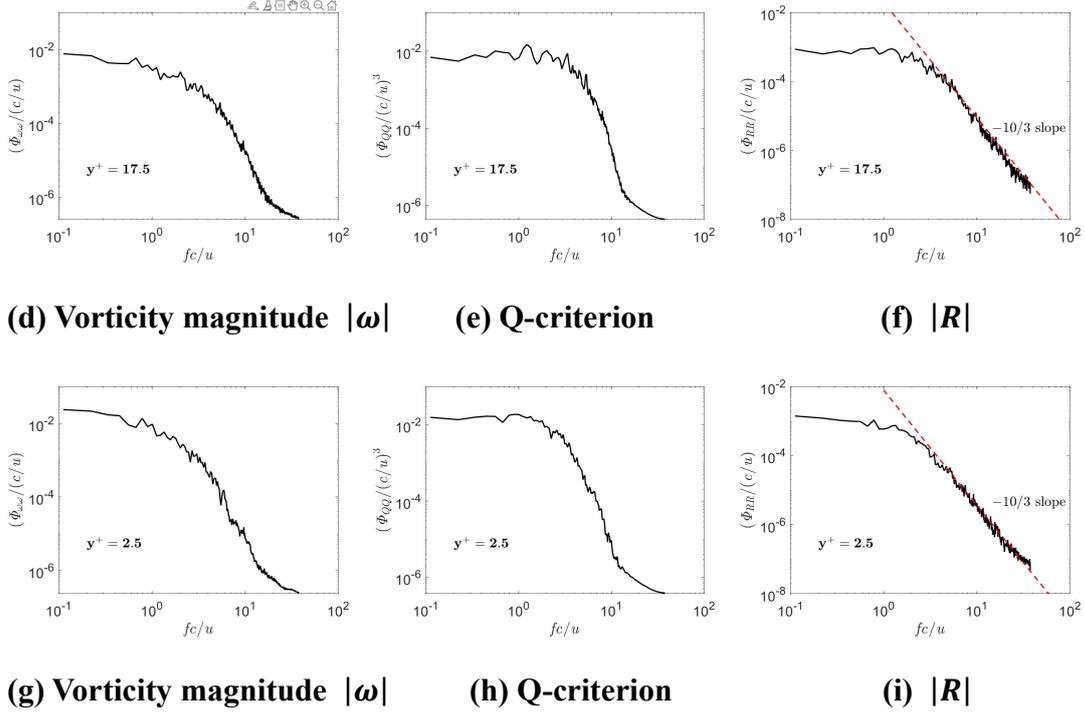

(d) Vorticity magnitude $|\omega|$     (e) Q-criterion     (f) $|R|$

(g) Vorticity magnitude $|\omega|$     (h) Q-criterion     (i) $|R|$

Fig. 5. Power spectrum density (PSD) of different vortex identification methods at various wall-normal locations ($y^+ = 2.5,\ 17.5\ \text{and}\ 70$)

The mean streamwise velocities obtained with different sub-grid scale models are plotted in wall units in Fig. 6, together with the classical laws $u^+ = y^+$ and $u^+ = (1/\kappa)\ln(y^+) + C$. Fig. 6(a) shows the velocity profiles for three grid resolutions using the present model. In the viscous sublayer, the velocity profiles obtained from all three grid resolutions agree well with the DNS results[31] and classical law $u^+ = y^+$. In the logarithmic region, the present model exhibits reduced errors with increasing grid resolution, demonstrating satisfactory grid convergence. Fig. 6(b) and Fig. 6(c) compares different sub-grid scale models on the medium and fine grids, respectively. 'SM' denotes the Smagorinsky subgrid-scale model and will be used in the following figures for simplicity. In the viscous sublayer, the SM model shows significant discrepancy from DNS results[31] and the classical law $u^+ = y^+$, which is attributed to its poor near-wall performance. However, this discrepancy decreases as the grid resolution increases. Incorporating the Van Driest damping function improves the near-wall performance of the SM model, particularly in the viscous sublayer. Compared with the SM model and the SM model with the Van Driest damping function, the present model and WALE perform better, yielding nearly identical results.

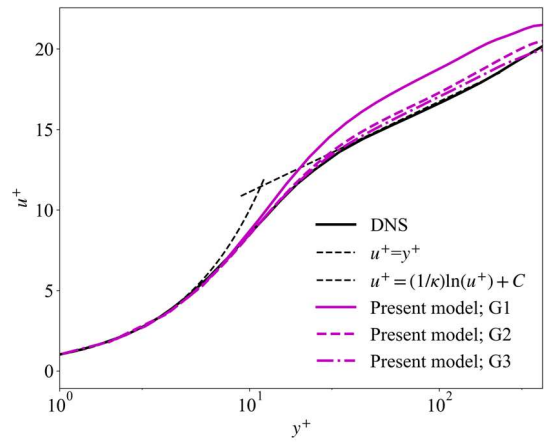

**(a) Present model for three grid resolutions**

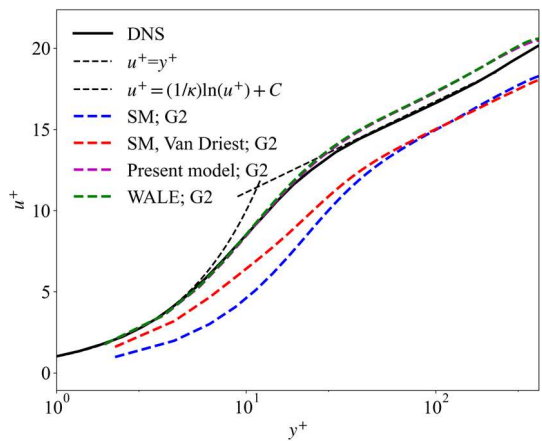 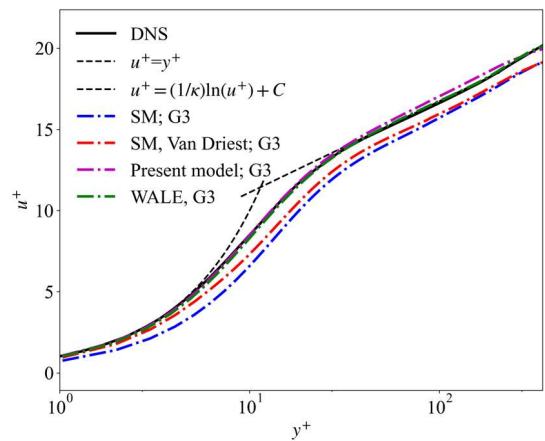

**(b) medium grid (G2)**          **(c) fine grid (G3)**

**Fig. 6. Mean streamwise velocities obtained with different sub-grid scale models**

From considerations of the behavior of the velocity fluctuations in the immediate vicinity of the wall, it is concluded that the eddy viscosity must grow at least as the cubic power of $y^+$. Fig. 7 presents the dimensionless eddy viscosity ($v_t/v$) near the wall. Fig. 7(a) shows the eddy viscosity for three grid resolutions using the present model. All eddy viscosities obtained from the present model approach zero near the wall, with an asymptotic trend proportional to $y$. Fig. 7(b) compares different sub-grid scale models for the medium grid. The Smagorinsky model (SM) displays high eddy viscosity near the wall，which is physically inaccurate. The addition of the Van Driest damping function enables the SM to exhibit improved near-wall characteristics, resulting in a vanishing tendency proportional to $y^2$. The WALE model exhibits the correct vanishing behavior, proportional to $y^3$, near the wall. The present model exhibits a vanishing tendency proportional to $y^2$, which is less accurate than that of WALE. Additionally, it can be observed that the values of $v_t/v$ from the present and WALE models fall within the range of $10^{-2}$ to $10^{-4}$, which is relatively small. Therefore, the present model can perform similarly to WALE in real engineering applications.

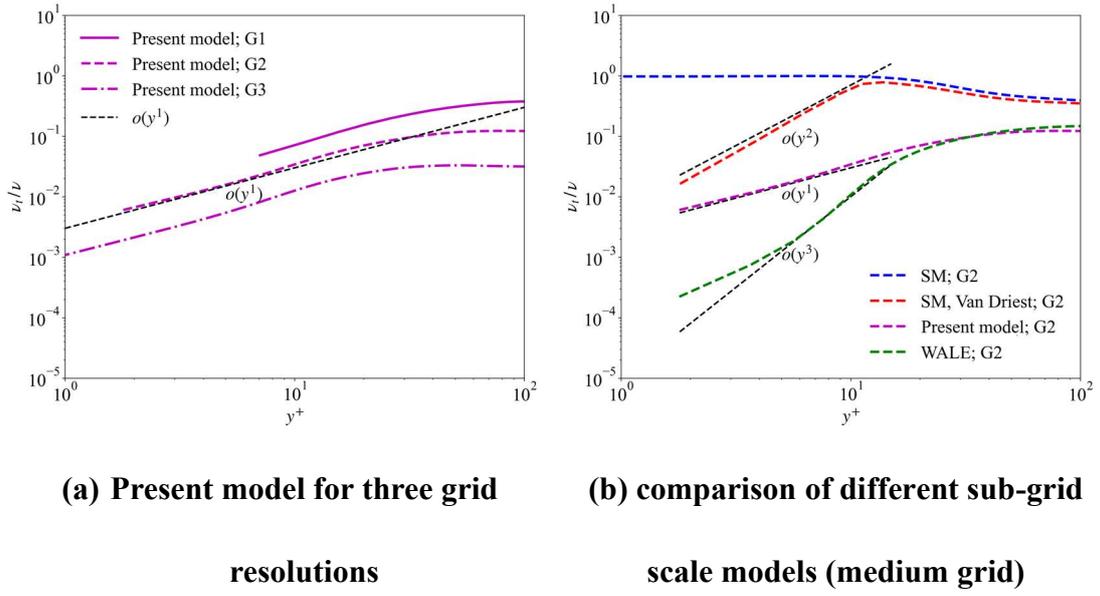

(a) Present model for three grid resolutions

(b) comparison of different sub-grid scale models (medium grid)

Fig. 7. Dimensionless eddy viscosity ($\nu_t/\nu$) near the wall

## 3.2. Periodic hill

The periodic hill is a classic case used to study separated and reattached flows. The computational domain consists of two hills, each with a height of $h = 0.028m$, separated by a distance of $9h$. The domain has a height of $3.035h$ and a spanwise extent of $4.5h$. The upper boundary is a flat wall, while the lower boundary is a curved channel wall. Periodic boundary conditions are applied in streamwise and spanwise direction and non-slip conditions at the lower and upper wall. The flow moves from left to right and is driven by a uniform body force. The Reynolds number, based on the hill height $h$ and the bulk velocity $U_b$ above the crest is 10595.

Throughout the paper, reference quantities for length, velocity and time are $h$, $u_b$ and $h/u_b$, respectively. All data presented are made dimensionless with these quantities. The time step was chosen so as to result in a maximum local CFL number of 0.3. After 5 flow-passing periods, $t_x = 9h/u_b$, mean quantities were collected over a period of $10t_x$ and were also averaged in the spanwise direction. The suitability of this integration period was checked by investigating changes in the statistics at intermediate times. Average quantities are denoted with angular brackets, and fluctuations with respect to the mean by a prime.

Table 2 reports the computational parameters for the LES cases, where $N_x$, $N_y$, and $N_z$ denote the number of grid points in each direction, and $\Delta x$, $\Delta y$, and $\Delta z$ represent the corresponding grid spacings. Three sets of grids are used to assess grid convergence, with $\Delta x+$ decreasing from approximately 58.60 to 30.61 while maintaining an aspect ratio of $AR = \Delta x/\Delta z = 1.8$. Fig. 8 shows computational domain and three grids for periodic hill.

**Table 2: Computational grids for periodic hill**

| Mesh sets | $(N_x \times N_y \times N_z)$ | $\Delta x^+$ | $\Delta y^+$ | $\Delta z^+$ | Total |
|---|---|---|---|---|---|
| Coarse (G1) | $79 \times 67 \times 71$ | 58.60 | 0.97 | 32.59 | 0.38M |
| Medium (G2) | $119 \times 100 \times 107$ | 45.36 | 0.75 | 25.20 | 1.27M |
| Fine (G3) | $178 \times 150 \times 160$ | 30.61 | 0.51 | 17.14 | 4.20M |

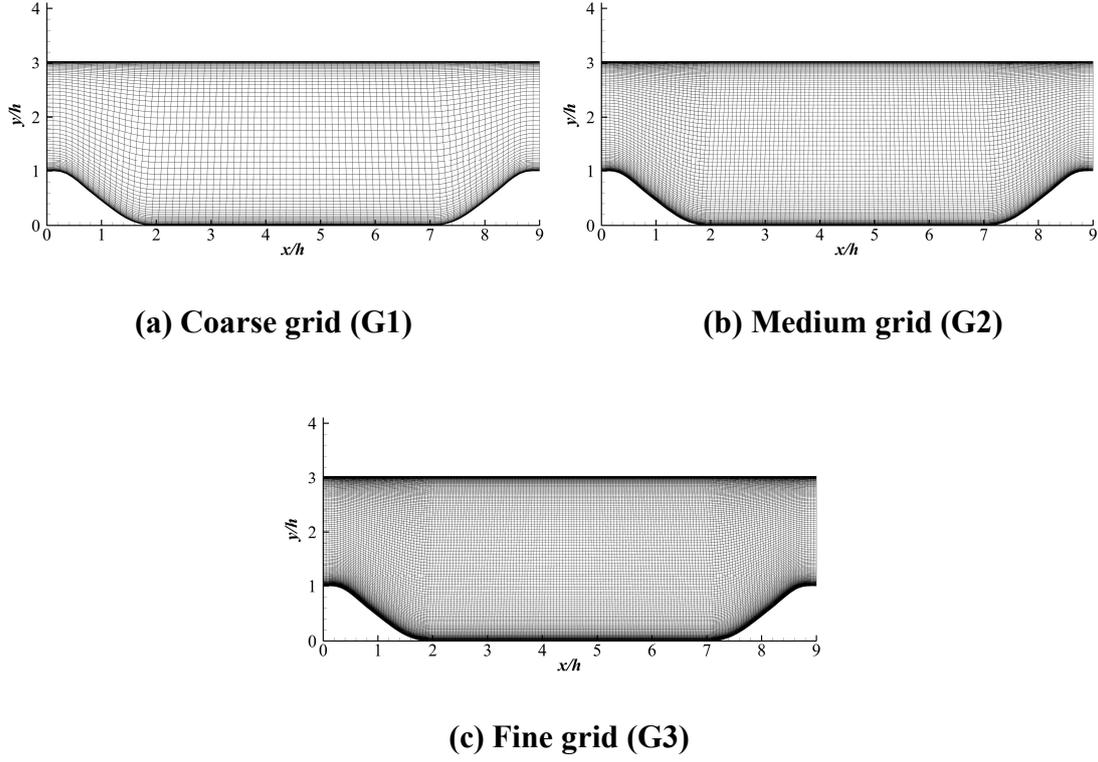

(a) Coarse grid (G1)　　(b) Medium grid (G2)

(c) Fine grid (G3)

**Fig. 8. Computational domain and three grids for periodic hill**

### 3.2.1 Results for time-averaged quantities

Fig. 9 shows skin friction coefficients of SM, SM (Van Driest), WALE, and the present models under different grids. The DNS results of Krank et al.[32] are used to validate the present simulations. All models, except for SM, can accurately predict the $C_f$ peak location. In Fig. 9(a), the skin friction coefficient exhibits significant variations in the separation region as grid resolution increases, whereas in Figs. 9(b)–(d), this effect is less pronounced. According to the experimental results reported by Ch. Rapp et al.[33], the reattachment point was measured at $x/h = 4.21$. For SM, the reattachment location shifts from $x/h = 5.3$ to $4.9$ and $4.8$ as the grid is refined from G1 to G3, with corresponding errors of 25.9%, 16.4% and 14.0%. In contrast, other models exhibit negligible changes in reattachment location. The reattachment points for the SM (Van Driest), WALE, and the present models are $x/h = 4.9$, 4.5, and 4.5, with errors of 16.4%, 6.9%, and 6.9%, respectively. In practical computations, the SM model is typically combined with the Van Driest damping function. Therefore, in the following study, the original SM model will no longer be considered for comparison.

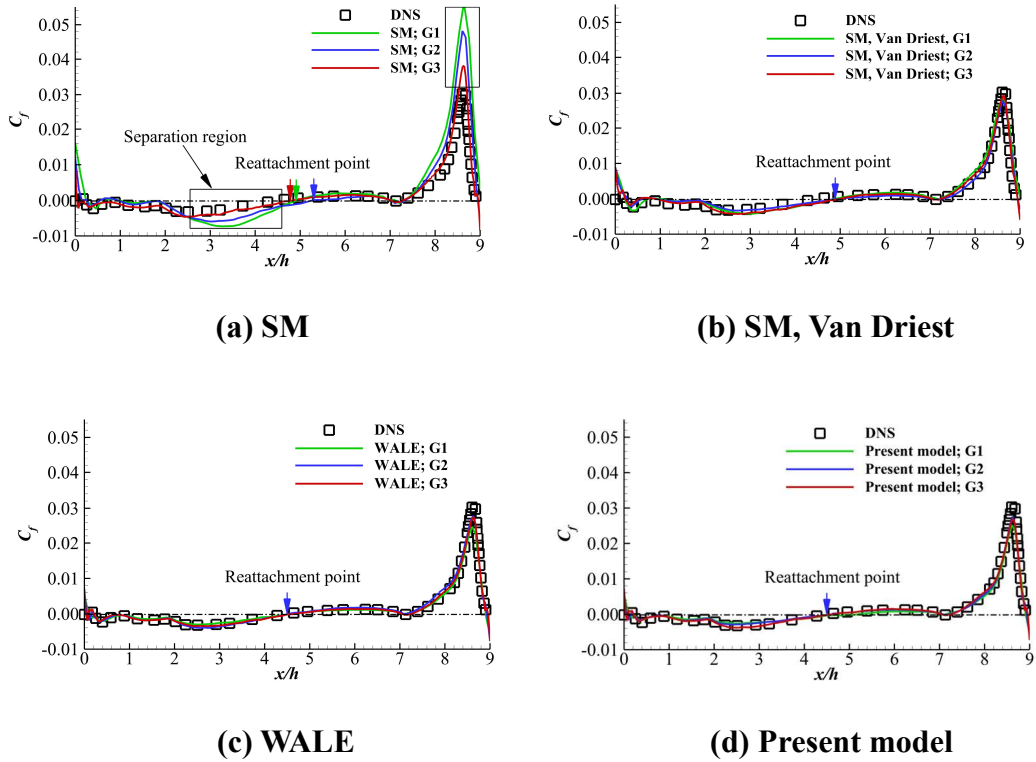

**Fig. 9. Skin friction coefficients $C_f$ of SM (Van Driest), WALE, and present model on different grids**

Fig. 10 shows the time-averaged separation and eddy viscosity using different sub-grid scale models. The reattachment locations are consistent with that shown in Fig. 9. The eddy viscosity in the separated shear layer varies across the different sub-grid scale models. The magnitudes follow the order: Smagorinsky with Van Driest model > WALE model > present model. The difference in separation caused by variations in eddy viscosity is not significant, particularly between the WALE and the present model. This is because eddy viscosity remains relatively small compared to kinematic viscosity, which dominates in most regions.

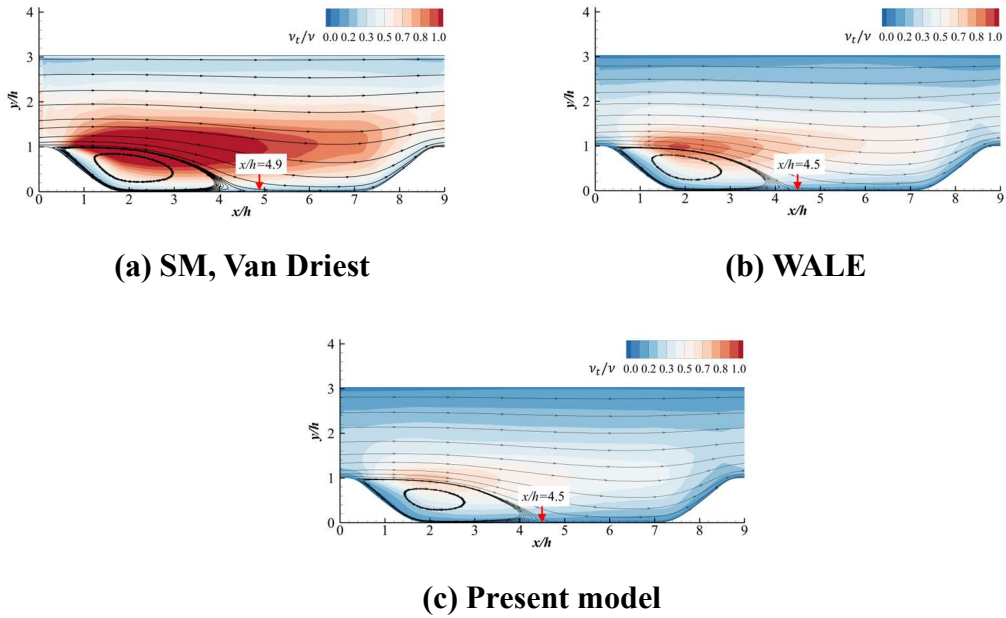

**Fig. 10. Time-averaged separation and eddy viscosity using different sub-grid scale models.**

The periodic hill case in this study contains several key flow regions that require special attention. Fig. 11 highlights four different streamwise locations: $x/h = 0.5$ (starting of separation), $x/h = 2$ (separation core), $x/h = 4$ (reattachment point), and $x/h = 6$ (fully reattached flow). These key positions will be analyzed in the following sections.

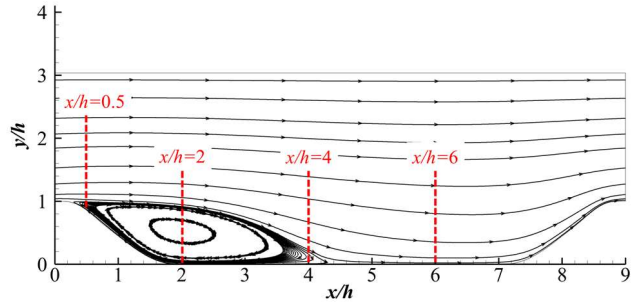

**Fig. 11.  Four different streamwise locations in periodic hill flow**

Fig. 12 presents the ratio of the eddy viscosity and kinematic viscosity ($v_t/v$) at four streamwise locations ($x/h = 0.5, 2, 4, 6$). At $x/h = 0.5$, the eddy viscosity from the present model gradually vanishes to zero near the wall, similar to the WALE model. The Smagorinsky model employs the Van Driest damping function to ensure zero values at the wall, but its predicted eddy viscosity remains significantly higher. At other streamwise locations, similar phenomena can be observed. Compared to channel flow, the periodic hill case is more complex due to curved geometry and flow separation. Nevertheless, the present model naturally vanishes eddy viscosity to zero at the wall.

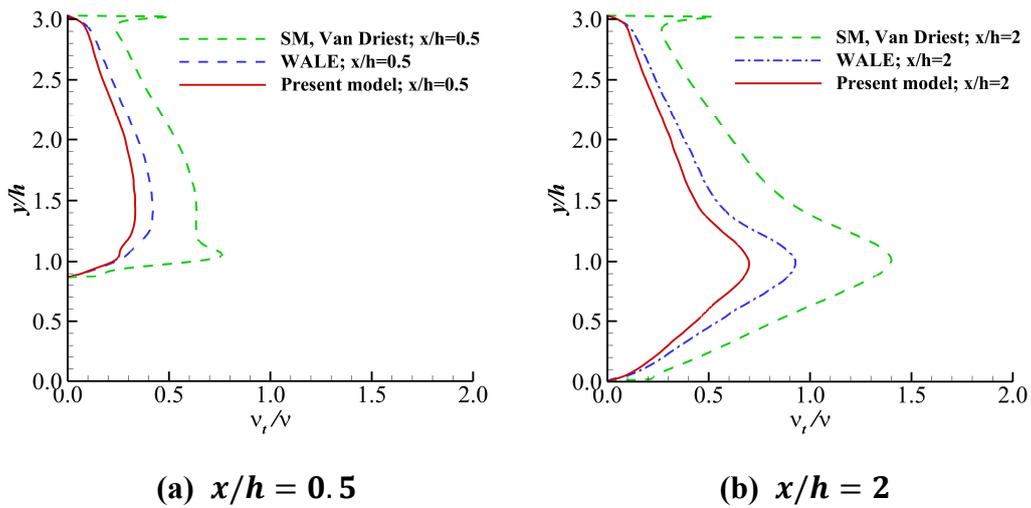

(a)  $x/h = 0.5$       (b)  $x/h = 2$

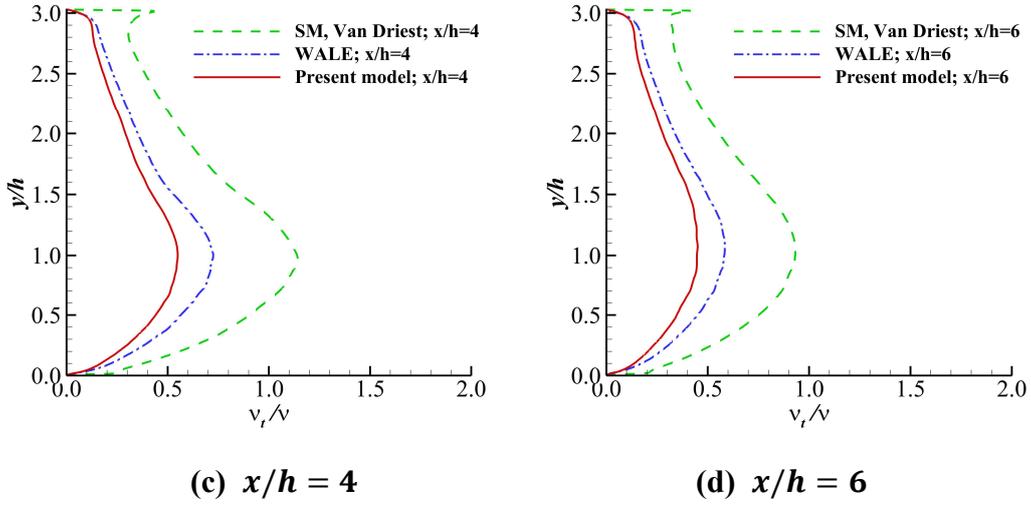

(c) $x/h = 4$  (d) $x/h = 6$

**Fig. 12. Ratio of the eddy viscosity and kinematic viscosity ($v_t/v$) at four streamwise locations. (a) $x/h = 0.5$, (b) $y/h = 2$, (c) $y/h = 4$, (d) $y/h = 6$**

Fig.13 presents the time-averaged streamwise velocity $\langle v \rangle$ at different locations predicted by different sub-grid scale models. The experimental results of Breuer et al.[34] at streamwise positions $x/h = 0.5$，2，4, and 6 are used to validate the present simulations. At the onset of separation ($x/h = 0.5$) and separation core ($x/h = 2$), the three models exhibit minimal differences. At the reattachment location ($x/h = 4$), the present model and WALE outperform the Smagorinsky model. At fully reattached location ($x/h = 6$), the Smagorinsky model remains the least accurate, while the present model and WALE model show some differences.

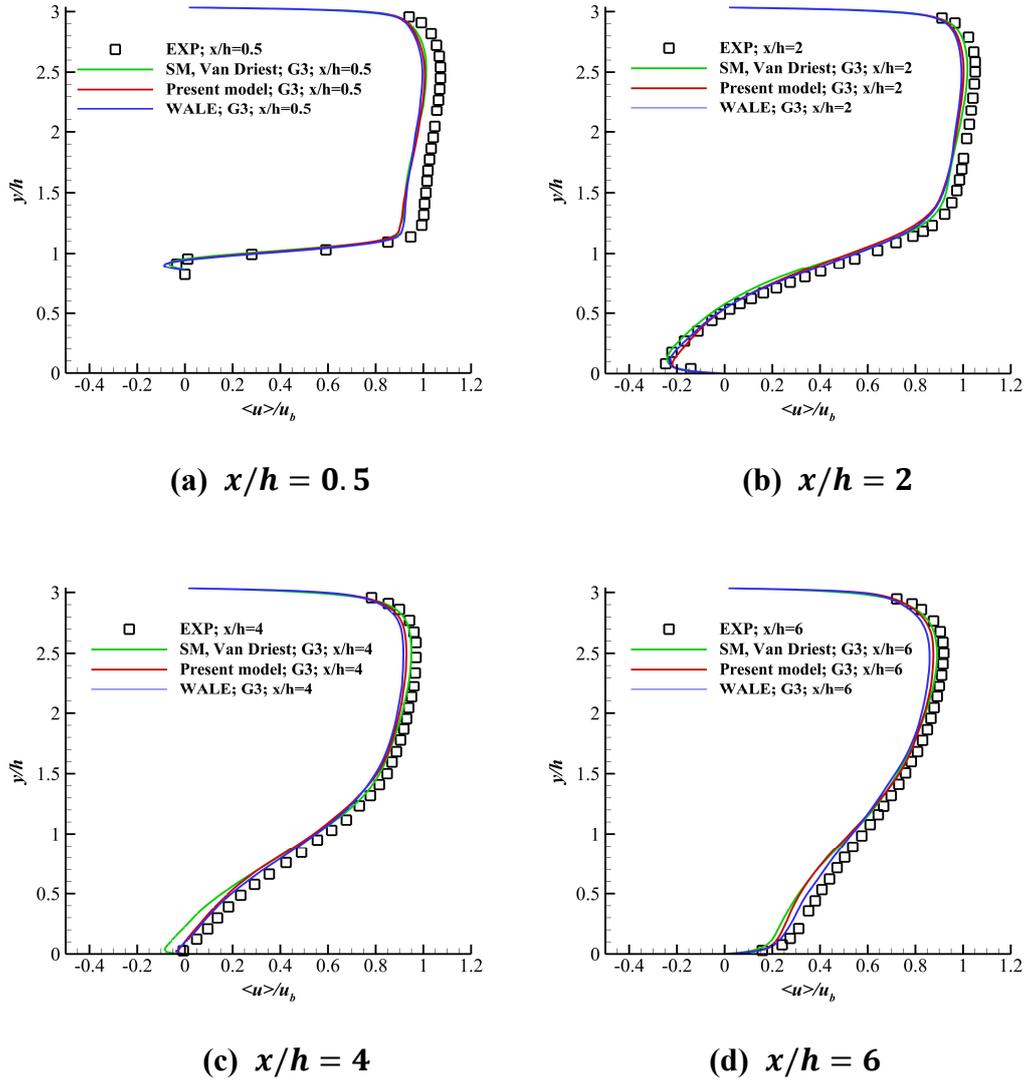

**(a)** $x/h = 0.5$  **(b)** $x/h = 2$

**(c)** $x/h = 4$  **(d)** $x/h = 6$

**Fig. 13. Time-averaged velocity $\langle u \rangle$ at different locations predicted by different models**

Fig. 14 shows the time-averaged velocity $\langle v \rangle$ at different streamwise locations predicted by various models. The experimental results of Breuer et al. [34] at streamwise positions $x/h = 0.5$，2，4, and 6 are used to validate the present simulations. At the initial separation location ($x/h = 0.5$) and the separation core ($x/h = 2$), the Smagorinsky model exhibits significantly larger errors compared to the other two models. The present model outperforms the WALE model at the separation core. At $x/h = 4$ and $x/h = 6$, all three models show minimal differences and closely match the experimental results.

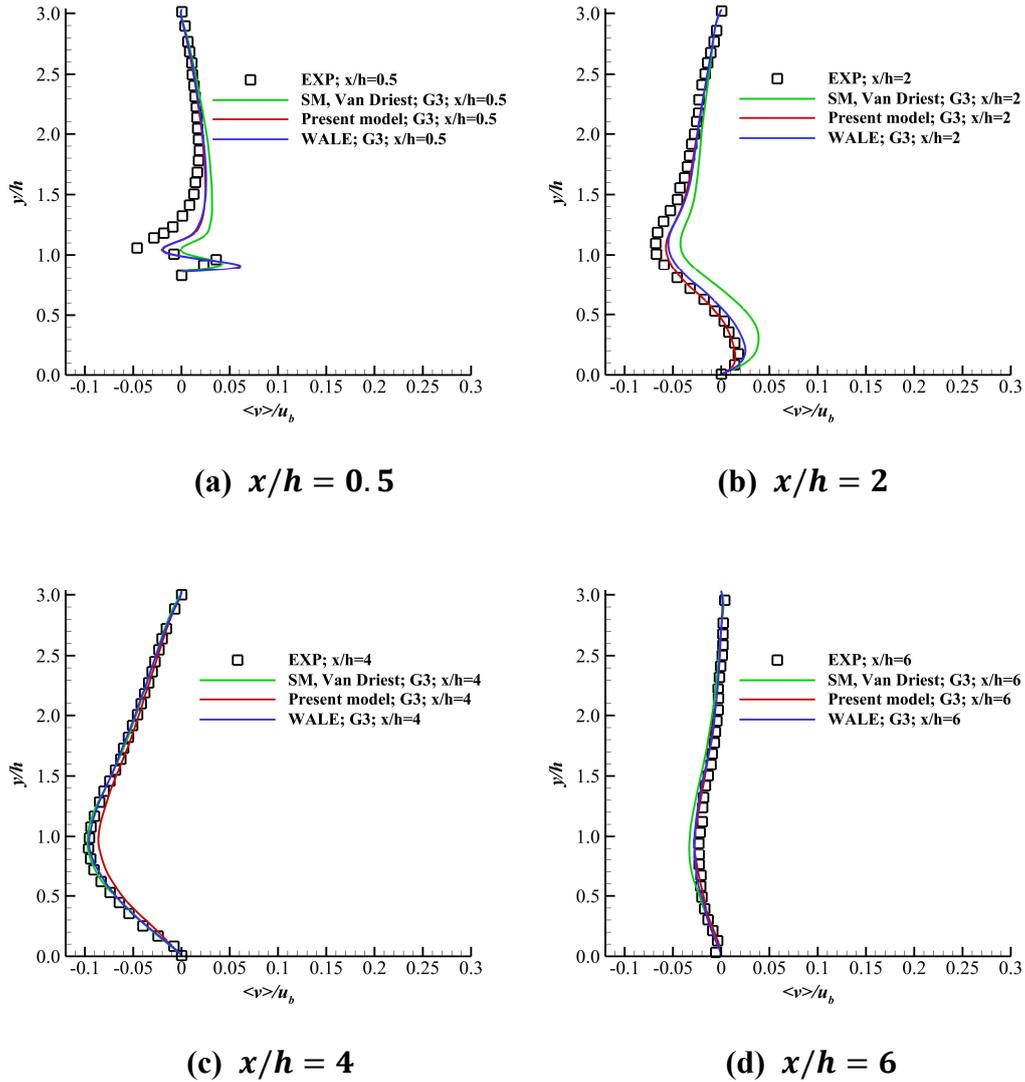

**Fig. 14. Time-averaged velocity $\langle v \rangle$ at different streamwise locations predicted by various models**

Fig. 15 shows comparison of time-averaged velocity under medium grid (G2) and fine grid (G3) at the vortex core streamwise location ($x/h = 2$). The results of mean velocity $\langle u \rangle/u_b$ (in Fig. 15(a)) shows that all three models agree well with the experimental results, with minimal differences among them. However, the results of mean velocity $\langle v \rangle/u_b$ (in Fig. 15(b)) reveal noticeable discrepancies between the models. On both medium grid (G2) and fine grid (G3), the present model and WALE model outperform the Smagorinsky model. While the present model and WALE model exhibit similar performance on the medium grid, the present model demonstrates superior accuracy on the fine grid.

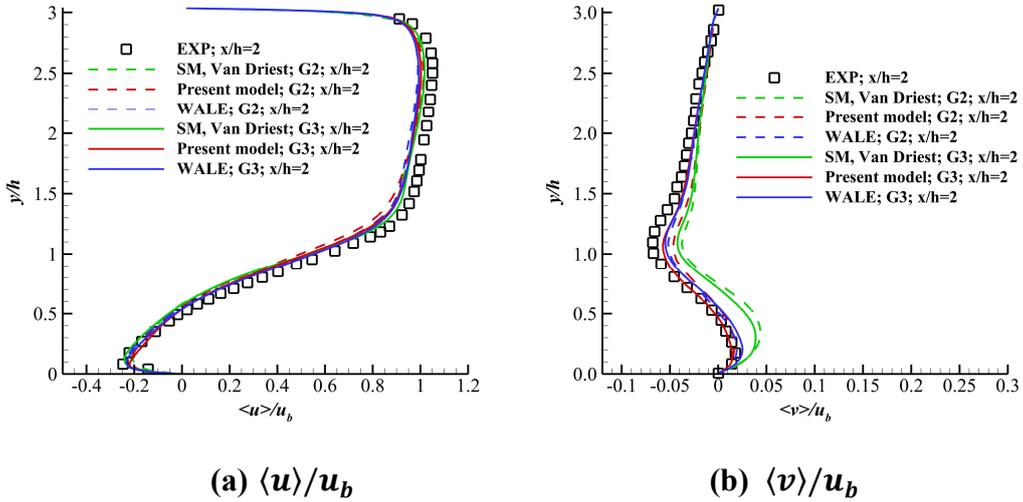

(a) $\langle u \rangle/u_b$        (b) $\langle v \rangle/u_b$

**Fig. 15. Comparison of time-averaged velocity under medium grid (G2) and fine grid (G3) at the vortex core streamwise location ($x/h = 2$)**

### 3.2.2 Results for instantaneous quantities

Fig. 16 shows the instantaneous flow and vortex based on the present model. The vortices are visualized using iso-surfaces of $|R| = 50$, and is contoured by dimensionless instantaneous velocity $u/u_b$. Time-averaged streamlines are provided as the background flow information. Separation starts from the crest and a large recirculation can be observed just behind the periodic hill.

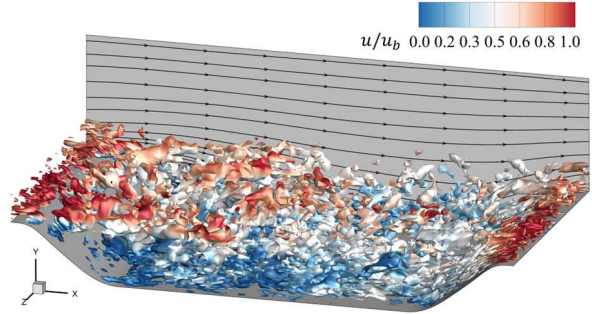

**Fig. 16. Instantaneous flow and vortices for $|R| = 50$**

Fig. 17 shows the Reynolds stress predicted by various models at different streamwise locations. The experimental results of Breuer et al. [34] at streamwise positions $x/h = 0.5$, 2, and 6 are used to validate the present simulations. At the onset and core of separation, the present model and the WALE model significantly outperform the Smagorinsky with Van Driest model. Fig.17 (b), (e), and (h) indicate that the present model outperforms the WALE model in the near-wall region. At the fully reattached flow location (*x/h*=6), the Reynolds stress predicted by the present model appears less developed compared to the other two models.

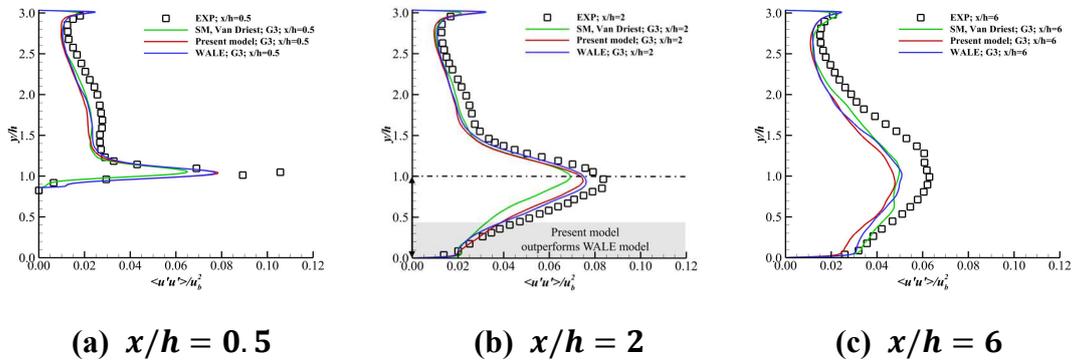

(a) $x/h = 0.5$   (b) $x/h = 2$   (c) $x/h = 6$

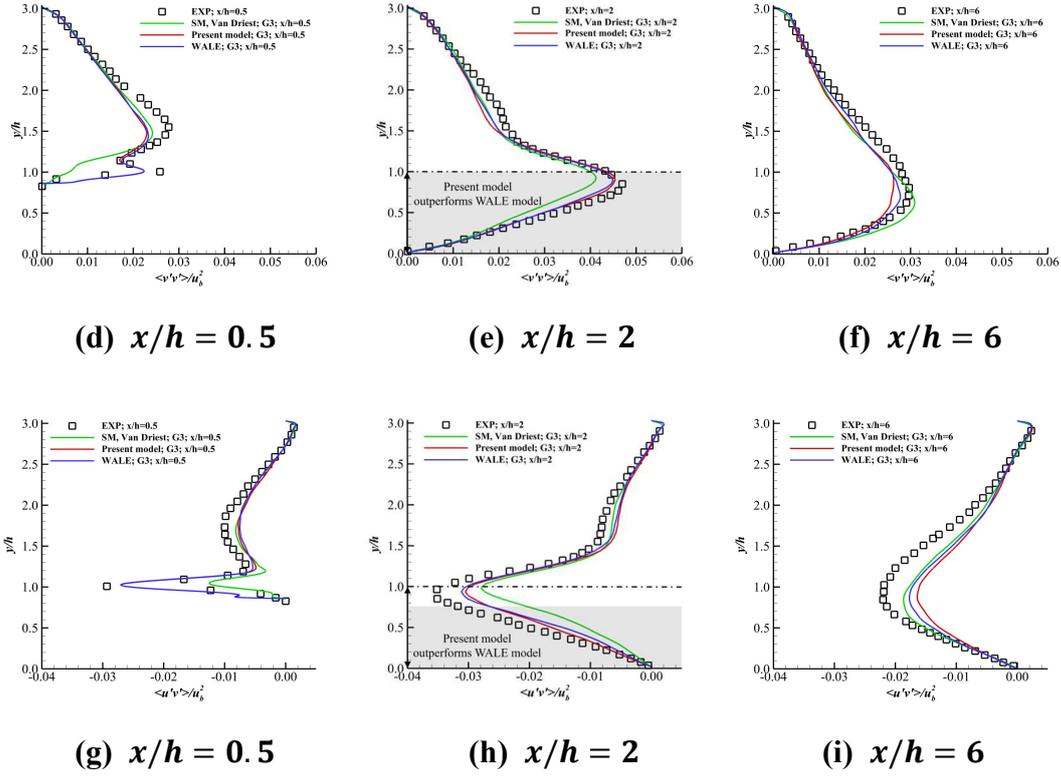

**(d)** $x/h = 0.5$  **(e)** $x/h = 2$  **(f)** $x/h = 6$

**(g)** $x/h = 0.5$  **(h)** $x/h = 2$  **(i)** $x/h = 6$

**Fig. 17. Reynolds stress predicted by various models at different streamwise locations**

Fig. 18 shows the comparison of Reynolds stress under medium grid (G2) and fine grid (G3) at the vortex core streamwise location ($x/h = 2$). The periodic hill case is primarily used to validate large separation, so the discussion should focus on the separation region ($y/h < 1.0$). Both the present model and the WALE model outperform the Smagorinsky model on medium and fine grids. On the medium grid, the present model provides more accurate predictions of $\langle u'u' \rangle/u_b^2$, $\langle v'v' \rangle/u_b^2$, $\langle u'v' \rangle/u_b^2$ components compared to the WALE model. On the fine grid, the present model exhibits superior performance in near-wall regions.

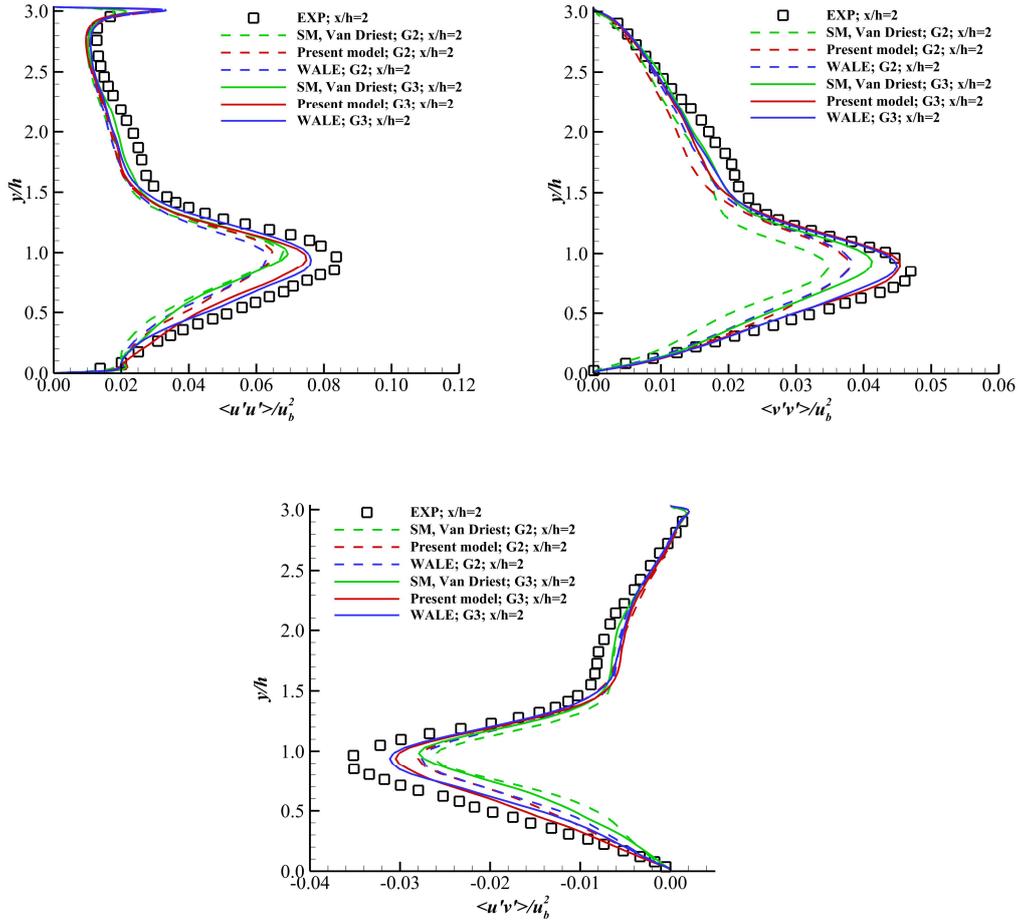

**Fig. 18. Comparison of Reynolds stress under medium grid (G2) and fine grid (G3) at the vortex core streamwise location ($x/h = 2$)**

## 4. Conclusions

This paper investigates the near-wall scaling and separation prediction capabilities of a rotation-based subgrid-scale stress model. Two classical cases, including channel flow and the periodic hill, are selected for investigation. The model's underlying mechanisms, spectral characteristics, and near-wall behavior are first studied. Then, its capability in predicting large-scale separation and Reynolds stresses is evaluated. The following conclusions can be drawn from this work:

(1) The characteristics of rigid rotation make it advantageous as a velocity scale introduced into subgrid stress models. $R$ represents rigid rotation, which naturally equals zero in the pure-shear viscous sublayer. Moreover, its PSD characteristics in the dissipative region follow a -10/3 law. The present model is vortex-based, which is both intuitive and mechanistic.

(2) In the near-wall region, the eddy viscosity of the present model naturally approaches zero, allowing it to perform similarly to the WALE model. In the channel flow case, it asymptotically decreases to zero following a $O(y)$ trend. Although its scaling order near the wall is not as $O(y^3)$ like in the WALE model, the dimensionless eddy viscosity of both this model and the WALE model lies between the range of $10^{-2}$ to $10^{-4}$. In practical calculations, both models exhibit essentially the same performance. The periodic hill case further demonstrates that the eddy viscosity becomes zero on complex geometries, such as curved surfaces.

(3) The present model can accurately predict large separation and performs better than other models in predicting Reynolds stress in the near-wall region. The Smagorinsky model shows significant deviations in predicting the reattachment point and surface friction coefficient. Even with the introduction of a damping function, the Smagorinsky model still exhibits noticeable deviations in predicting the reattachment point and the mean velocity within the separation region. In contrast, the present model performs better. Regarding Reynolds stress, the present model outperforms both the Smagorinsky model and the WALE model in the near-wall region with large separation.

Although this study demonstrates that the present model behaves better in near wall regions and can accurately predict large-scale separation. It also exhibits certain limitations. Its predictions in the reattachment region downstream of the separation are less accurate compared to the WALE model. Additionally, the model's coefficients may require adjustment for different flow conditions.

## Acknowledgments

This work is supported by the National Science Foundation, with Grant No. 2422573. The authors also thankful to TACC for providing computation resources.

## Author declarations

### Conflicts of Interest
The authors have no conflicts to disclose.

## Data availability

The data that support the findings of this study are available from the corresponding author upon reasonable request.